\documentclass[aip,amsmath,amssymb,preprint]{revtex4-1}
\usepackage[mathlines]{lineno}

\usepackage{graphicx}
\usepackage{dcolumn}
\usepackage{bm}
\usepackage[utf8]{inputenc}
\usepackage[T1]{fontenc}
\usepackage{mathptmx}
\usepackage{xcolor}
\usepackage{footnote}
\usepackage[symbol]{footmisc}
\definecolor{darkblu}{RGB}{46, 48, 146}
\usepackage[colorlinks=true,
            citecolor=darkblu,
            urlcolor=darkblu,
            linkcolor=black]{hyperref}

\begin{document}
\title{Enhanced Surface Superconductivity in Ba(Fe$_{0.95}$Co$_{0.05}$)$_2$As$_2$}

\author{Christopher T. Parzyck}
\affiliation{Department of Physics, Laboratory of Atomic and Solid State Physics, Cornell University, Ithaca, NY 14853, USA}
\author{Brendan D. Faeth}
\affiliation{Department of Physics, Laboratory of Atomic and Solid State Physics, Cornell University, Ithaca, NY 14853, USA}
\author{Gordon N. Tam}
\affiliation{Department of Physics, University of Florida, Gainesville, FL 32611, USA}
\author{Gregory R. Stewart}
\affiliation{Department of Physics, University of Florida, Gainesville, FL 32611, USA}
\author{Kyle M. Shen\footnote{Author to whom correspondence should be addressed: kmshen@cornell.edu}}
\affiliation{Department of Physics, Laboratory of Atomic and Solid State Physics, Cornell University, Ithaca, NY 14853, USA}

\date{\today}

\begin{abstract}
We present direct evidence for an enhanced superconducting $T_c$ on the surface of cleaved single crystals of Ba(Fe$_{0.95}$Co$_{0.05}$)$_2$As$_2$.  Transport measurements performed on samples cleaved in ultra high vacuum (UHV) show a significantly enhanced superconducting transition when compared to equivalent measurements performed in air. Deviations from the bulk resistivity appear at 21K, well above the 10K bulk $T_c$ of the underdoped compound. We demonstrate that the excess conductivity above the bulk $T_c$ can be controllably suppressed by application of potassium ions on the cleaved surface, indicating that the enhanced superconductivity is strongly localized to the sample surface.  Additionally, we find that the effects of the potassium surface dosing are strongly influenced by the presence of residual gas absorbates on the sample surface, which may prevent effective charge transfer from the potassium atoms to the FeAs plane. This is further support for the conclusion that the effects of the dosing (and enhanced superconductivity) are localized within a few layers of the surface.\\[24pt]

This article may be downloaded for personal use only. Any other use requires prior permission of the author and AIP Publishing. This article appeared in Applied Physics Letters and may be found at \href{https://aip.scitation.org/doi/full/10.1063/1.5133647}{https://aip.scitation.org/doi/full/10.1063/1.5133647}
\end{abstract}

\maketitle

The field of high temperature superconductivity, particularly in the cuprates and iron pnictides, has benefited greatly from complementary information from both bulk-sensitive thermodynamic (transport, calorimetry) and surface-sensitive spectroscopic (angle resolved photoemission, scanning tunneling microscopy) probes. However, care must be taken when connecting properties derived from surface sensitive techniques to their bulk counterparts.  For instance, symmetry breaking at the material/vacuum interface can lead to electronic reconstructions and charge transfer at the surface.  Such a `self doping' effect has been reported at the surface of cleaved single crystals of YBa$_2$Cu$_3$O$_{7-\delta}$ where the surface termination heavily influences the apparent doping observed in angle resolved photoemission (ARPES) experiments\cite{Zabolotnyy2007,Nakayama2007,Hossain2008}.

\begin{figure*}
\includegraphics[width=\textwidth]{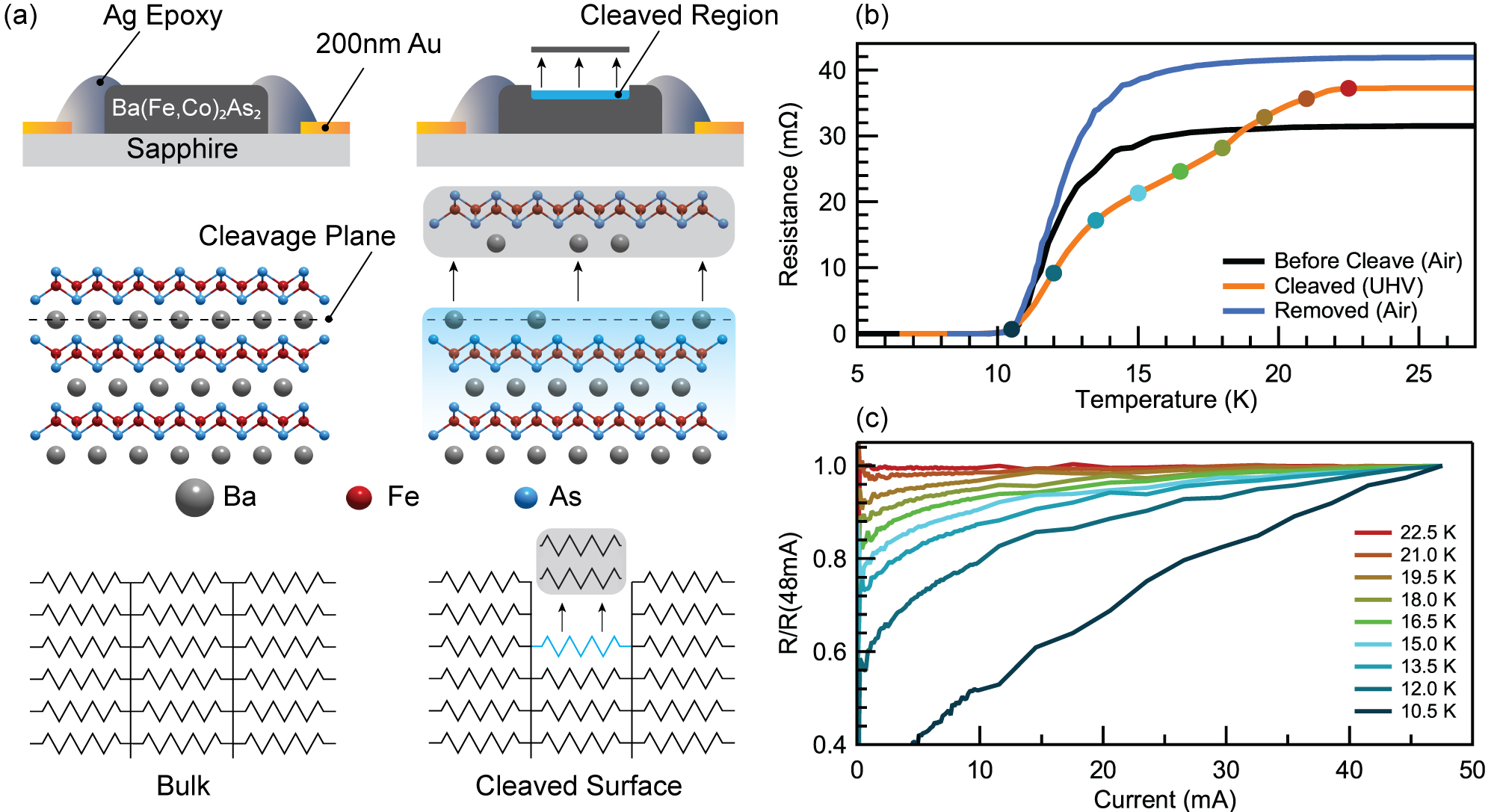}
\caption{\label{fig:cleaving} (a) Schematic of measurement.  A four point corner contact geometry is used to eliminate wiring and contact resistances from the measurement. However, the cleaved surface (teal) is unavoidably measured in parallel, and in series, with the uncleaved regions of the sample. (b) Four point resistance of sample before cleaving, after cleaving, and after exposure of cleaved sample to air. A measurement current of 50$\mu$A was used, at this value no noticeable change in the bulk $R(T)$ curve was observed with changing applied current.  (c) Measured resistance as a function of applied current, $R(I) = V(I)/I$, for the cleaved sample; deviations from a constant value indicate non-ohmic behavior. For clarity, each of the curves have been normalized to the resistance obtained at 48mA of applied current.}
\end{figure*}

In this letter, we present evidence for a modified superconducting state on the surface of cleaved single crystals of underdoped Ba(Fe,Co)$_2$As$_2$.  Due to the weaker bonds between the Ba layer and FeAs layer than between the Fe and As atoms themselves, single crystals of Ba(Fe,Co)$_2$As$_2$ tend to cleave at the Ba layer.  To maintain charge neutrality, half of the Ba ions should remain on each of the separated surfaces (Figure \ref{fig:cleaving}a). As a result the region adjacent to the cleaved surface, the selvedge\cite{woodruff2016}, differs both chemically and structurally from the bulk.  The remaining surface Ba atoms have been shown to form $2\times 1$, $\sqrt{2}\times\sqrt{2}$, and occasionally $1\times 1$ reconstructions\cite{Koepernik2012,Li2019a,Teague2011,Massee2009a,Massee2009,Zhang2010} depending on cleavage temperature.  Additionally, reconstructions typically vary across a single sample and may be separated by regions of disordered Ba adatoms or areas of As termination.  Low energy electron diffraction (LEED-IV) measurements performed on undoped samples cleaved at low temperature indicate structural distortions from the bulk lattice on the As terminated surfaces\cite{Nascimento2009}, and similar measurements on Ca(Fe,Co)$_2$As$_2$ show buckling of the Fe-As layer on $2\times 1$ reconstructed surfaces\cite{Li2014}. Either an uneven distribution of the Ba adatoms, or structural distortions in the surface lattice (or a combination of the two) may lead to a modified doping and changes in the superconductivity in the selvedge.  Indeed, peak shifts in x-ray photoemission spectroscopy indicate a different electrostatic environment associated with the atoms on the surface \cite{Perisse2011}. However, the specific effect this has on the electronic transport and superconductivity is not completely clear.

In an effort to directly connect the aforementioned effects seen in surface probes to the transport properties, we have performed a series of \textit{in situ} transport measurements on cleaved samples of underdoped Ba(Fe$_{1-x}$,Co$_x$)$_2$As$_2$, $x=0.05$.  Single crystal samples were grown using the flux method and annealed using the procedure described in Ref. \onlinecite{Tam2013}.  Square samples with dimensions approximately $2\times 2 \times 0.5$ mm were selected and mounted on sapphire plates using an insulating, UHV compatible epoxy. Contacts between the sample corners and pre-patterned, 200 nm thick, Au pads were made using a conducting Ag-based epoxy. Four point resistance measurements were performed using a Keithley 6221 current source and Keithley 2182A voltmeter in a delta mode configuration.  With an applied current of 50 $\mu$A, pulse width of 19 ms, and repetition rate of 27 Hz, no heating effects, or changes in the bulk superconducting transition, were observed. Baseline measurements of the underdoped samples were first taken in air and showed a slow rollover with 10 and 25 percent drops in resistance occurring at $T_c^{90\%}=14.4$K and $T_c^{75\%}=13.2$K.  The samples were then cleaved in UHV ($P<7\times 10^{-11}$ torr) at room temperature and transferred into the measurement chamber ($P<9\times 10^{-11}$ torr) where electrical contact was made to the Au pads with a set of spring loaded pins. As can be seen in the transport curves (Figure \ref{fig:cleaving}b) the normal state resistance of the sample increases by 18$\%$ when cleaved, as is expected due to the reduction in thickness caused by the cleaving process. More importantly, the shape of the transition is dramatically altered; the resistance decreases rapidly below 21K, with increases in $T_c^{90\%}$ and $T_c^{75\%}$ to 19.7K and 17.5K, respectively -- a change of over 4K from the bulk values.  The zero resistance $T_c=10.4$K and the resistance upturn associated with the spin density wave at $T_{SDW}=82$K (not plotted here) remain the same before and after cleaving.  Finally, when the sample is removed from vacuum and exposed to air the anomalous drop in resistance below 21K disappears, returning to the bulk behavior measured before cleaving.  

We propose that this decrease in resistance in the UHV cleaved samples can be attributed to an enhanced superconducting $T_c$ at the sample surface. To investigate the origin of the excess conductivity, we have measured the V(I) characteristics of the cleaved samples in this regime.  Pulsed V(I) measurements were conducted using the same Keithley current source/voltmeter previously mentioned. Using a 500 $\mu s$ pulse width, and a 0.5\% duty cycle the average power was sufficiently low that no heating effects were observed at any applied current.  Because the changes in the voltage characteristics are small and deviate only at low currents, we plot (Figure \ref{fig:cleaving}c) the sample resistance ($V(I)/I$) as a function of applied current ($I$).  In this configuration, deviation from a \emph{linear} (ohmic) voltage response are shown as a deviation from a \emph{constant} resistance.  A noticeable change in the low current behavior occurs between 19.5K and 21K, well above the bulk $T_c$ and coincident with the sharp change in resistivity.  This onset of a non-linear voltage response, and its suppression at high applied currents, is highly suggestive that the corresponding drop in resistance is due to part of the sample surface undergoing a superconducting transition.

Presuming this effect arises from the sample surface, it should respond to modification of surface electronic structure; such modification can be accomplished \textit{in situ} by means of alkali metal surface dosing.  The deposition of small quantities of K, Rb, or Cs on a pristine surface can introduce electrons to the topmost layers, effectively altering the doping in the selvedge.  Consequently, this technique has been used in conjunction with surface sensitive spectroscopic probes, such as ARPES and STM, to monitor the doping evolution of the surface electronic structure in a wide variety of quantum materials systems including cuprates\cite{Hossain2008,Fournier2010}, iridates\cite{Schindler2014,Kim2016a}, iron based superconductors\cite{Seo2016,Kyung2016,Zhang2016,Ren2017}, topological insulators \cite{Zhu2011}, and quantum well states \cite{Carlsson1997,Breitholtz2007}.  In surface sensitive investigations involving alkali metal surface dosing, it is typically believed that the donated electrons are primarily confined to within the uppermost unit cells (i.e. 1-2) closest to the surface, although this depth likely depends strongly on the Thomas-Fermi screening length.  In this study, we utilize potassium dosing and \textit{in situ} transport measurements to demonstrate that the secondary resistive transition observed in our cleaved samples is localized to the surface, further supporting the hypothesis that it is indeed due to an enhancement of the superconductivity there.  

\begin{figure}
\includegraphics[width=0.5\textwidth]{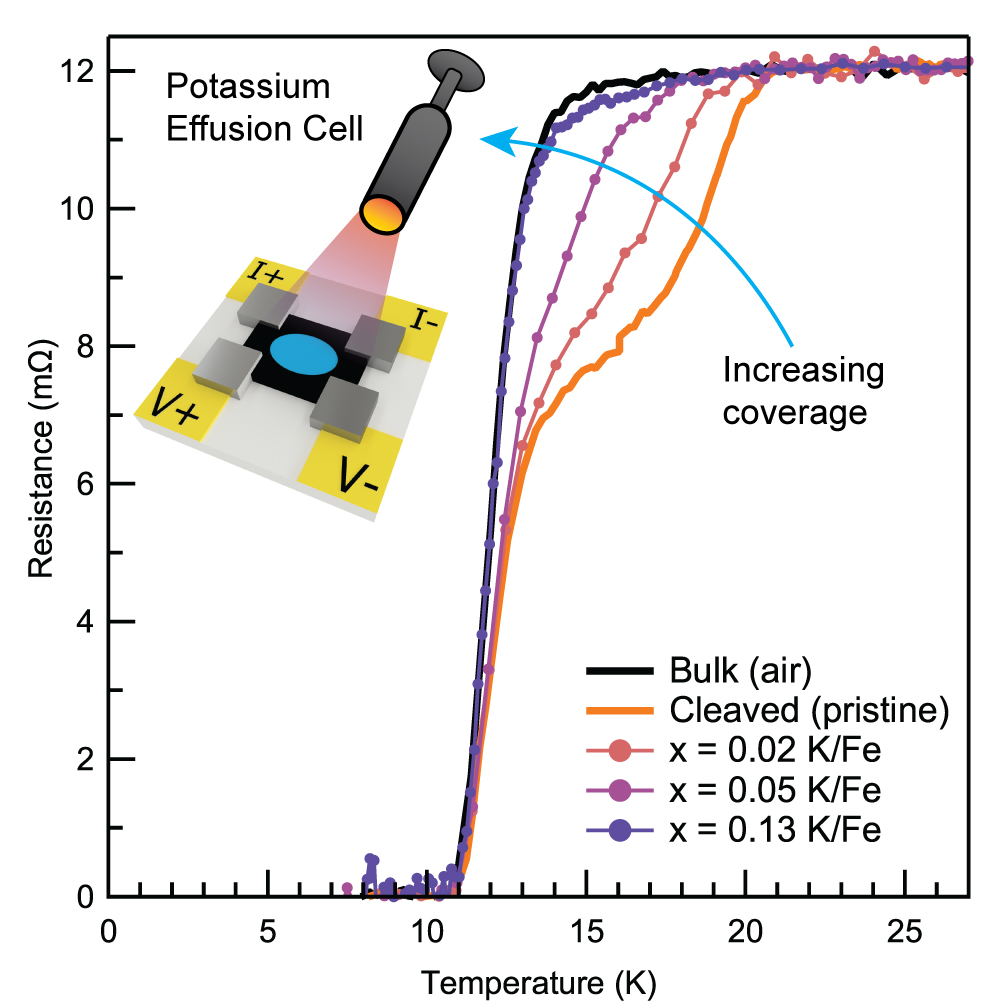}
\caption{\label{fig:dosing} Effects of potassium surface dosing on the surface contribution to the superconducting transition.  The sample surface was dosed using a shuttered molten potassium source and the sample temperature was kept between 7K and 35K during the sequence; no evidence of potassium desorption was observed in this temperature range.  In all measured samples a monotonic suppression of the secondary transition was observed with increasing potassium coverage. Quoted nominal surface coverages, $x$, are based on flux measurements of the potassium source by quartz crystal microbalance.}
\end{figure}

A shuttered effusion cell with a molten potassium source (flux calibrated using a quartz crystal microbalance) was used to controllably and repeatably apply fractions of a monolayer of potassium to the cleaved sample surfaces.  Between each dose the resistance between 7K and 35K was remeasured, with results plotted in Figure \ref{fig:dosing}. As the amount of deposited potassium was increased, the anomalous dip in resistivity was suppressed, saturating at a nominal surface coverage of 0.13 K ions per Fe atom.  The response of the transition to surface doping is consistent with the hypothesis that the additional conductivity is localized to the surface of the material, as carriers donated from the adsorbed K ions likely only permeate a few unit cells into the material\cite{Zheng2016}.  Assuming an approximate sample thickness of 500 $\mu$m and an active surface layer thickness of 1 nm, the 30\% drop in measured resistance corresponds to a drop in surface resistance on the order of $10^{5}$, which is a lower bound that neglects series resistance effects from the uncleaved regions.  Such a drastic change in conductivity implies the presence of superconductivity well above the bulk $T_c$ in the as cleaved surface, as substitutional doping studies indicate that changes in the low temperature normal state resistance are confined to within an order of magnitude across a wide range of both hole and electron doping\cite{Olariu2011,Shen2011}.  Although we consider doping due to charge transfer from the potassium ions as the most likely source of suppression of the surface superconductivity, it is also possible that `re-filling' some of the Ba vacancies left by the cleave may also cause subtle structural changes at the surface that influence the low energy physics.  This is another avenue by which the superconductivity may be suppressed, given the reported sensitivity of the iron pnictide superconductors to the Fe-Pn bond parameters\cite{Vildosola2008,Kuroki2009,Drotziger2010,Tomic2012}.
 
\begin{figure*}
\includegraphics[width=\textwidth]{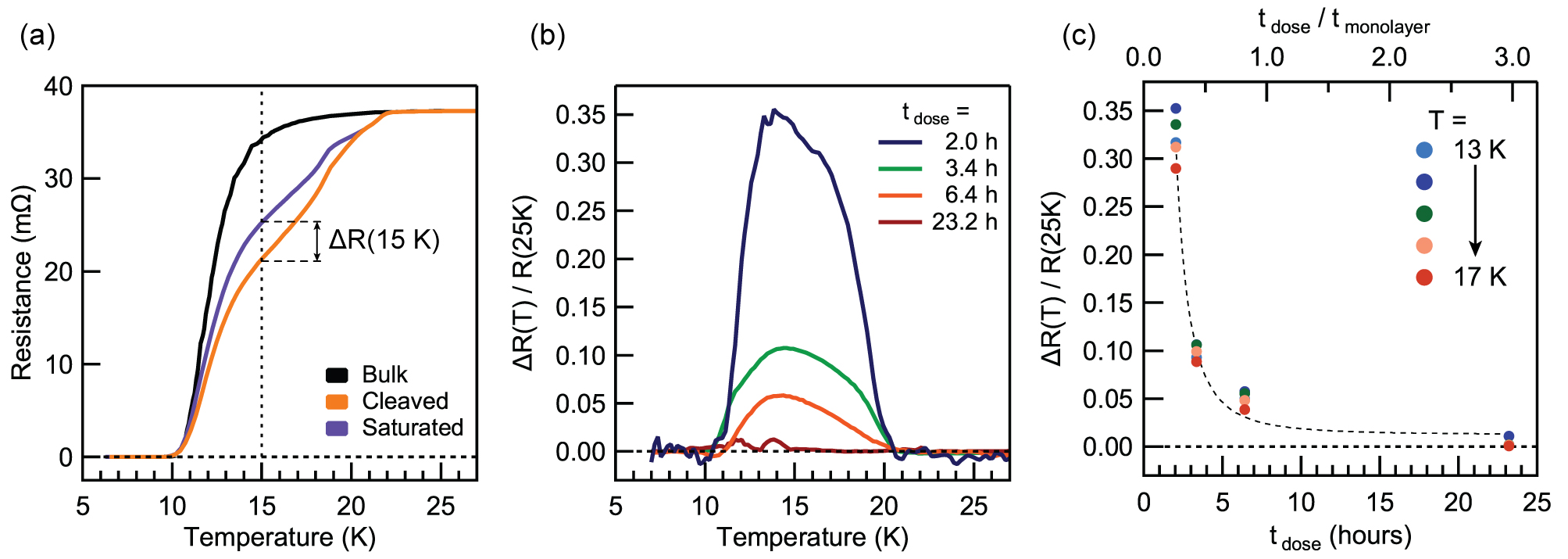}
\caption{\label{fig:aging} Effects of residual gas adsorbtion on potassium surface doping results.  (a) Waiting a time of t$_{\,\textrm{dose}}= 3.4$ hours between cleaving and dosing, the saturated resistance after the dosing sequence remains significantly below that of the bulk. (b) Difference in sample resistance, $\Delta$R(T), before and after potassium dosing. Curves are normalized to the resistance at 25K to account for sample-to-sample variations in contact geometry and thickness. (c)  As the delay time between the initial cleave and start of the dosing series (t$_{\,\textrm{dose}}$) increases beyond a few hours the effects of the potassium surface dosing are minimized.  This time scale correlates with the formation time for a monolayer of residual gasses on the sample surface (t$_{\,\textrm{monolayer}}$), given the chamber background pressure of $9\times 10^{-11}$ torr.  Dashed line is a guide to the eye.}
\end{figure*}

A final piece of evidence which points towards the surface-derived nature of the enhanced superconductivity is the sensitivity of potassium dosing effects to residual gas adsorbates on the sample surface. By varying the delay time between cleaving the sample and dosing the surface, we have examined how the residual gas adsorbates modify the effects of potassium deposition on a series of samples which displayed the double transition pictured in Figure \ref{fig:dosing}. As this delay time approaches the formation time for a monolayer of residual gas adsorbates, the efficacy of the potassium dosing is substantially reduced (Figure \ref{fig:aging}). After even a few hours in vacuum we observe that the resistance after the dosing sequence saturates at a value below that of the bulk.  As a layer of adsorbed gas molecules builds up on the sample surface, it likely impedes effective charge transfer from the potassium ions to the top FeAs layer, which is consistent with effects of the dosing (and enhanced superconductivity) being localized to the top few layers of the sample.

Using transport measurements performed in UHV, we have demonstrated that the surface of cleaved Ba(Fe$_{0.95}$Co$_{0.05}$)$_2$As$_2$ hosts a modified superconducting state with an onset $T_c$ at least 4K greater then measured in the bulk of the same samples.  This enhanced $T_c$ is observed only in vacuum and likely stems from electronic or structural reconstructions induced by the cleave.  Furthermore, the enhanced superconductivity can be regulated by the addition of additional electrons donated from alkali metals deposited on the sample surface.  This indicates the enhancement is confined to a thin selvedge layer, however determination of the precise structure, thickness, and doping of this region remain a challenge.  The presence of such a modified state at the cleaved surface illustrates that care must be taken when comparing results from surface sensitive probes of the electronic structure to their bulk counterparts and the usefulness of \textit{in situ} transport experiments in understanding the transport properties of both the surface and the bulk of materials.


This work was supported through the National Science Foundation [Platform for the Accelerated Realization, Analysis, and Discovery of Interface Materials (PARADIM)] under Cooperative Agreement No. DMR-1539918, NSF DMR-1709255, and the Air Force Office of Scientific Research Grant No. FA9550-15-1-0474. This research was also funded in part by the Gordon and Betty Moore Foundation's EPiQS Initiative through Grant No. GBMF3850 to Cornell University. This work made use of the Cornell Center for Materials Research Shared Facilities which are supported through the NSF MRSEC program (DMR-1719875).  G.N. Tam and G.R. Stewart acknowledge support through US Department of Energy, Basic Energy Sciences, Contract no. DE-FG02-86ER45268, B.D. Faeth acknowledges support from the NSF Graduate Research Fellowship under Grant No. DGE-1650441, and C.T. Parzyck acknowledges support from the Center for Bright Beams, NSF award PHY-1549132. 

\bibliography{./library}
\end{document}